\documentclass[conference]{IEEEtran}
\IEEEoverridecommandlockouts
\usepackage{cite}
\usepackage{amsmath,amssymb,amsfonts}
\usepackage{algorithmic}
\usepackage{graphicx}
\usepackage{textcomp}
\usepackage{comment}
\newtheorem{assumption}{Assumption}
\newtheorem{definition}{Definition}
\newtheorem{lemma}{Lemma}
\newtheorem{theorem}{Theorem}
\usepackage{xcolor}
\usepackage{times}
\usepackage{amsfonts,amssymb}

\renewcommand{\raggedright}{\leftskip=0pt \rightskip=0pt plus 0cm}
\usepackage{soul}
\usepackage{url}
\usepackage[hidelinks]{hyperref}
\usepackage[utf8]{inputenc}
\usepackage[small]{caption}
\usepackage{graphicx}
\usepackage{amsmath}
\usepackage{booktabs}
\usepackage{comment}
\usepackage{amsmath,lipsum}
\usepackage{cuted}%%\stripsep-3pt
 \usepackage{algorithm}
\usepackage{amsmath}

\usepackage{amsfonts}
\usepackage{bbm}
\usepackage{dsfont}
\usepackage{diagbox} % 加载宏包
\usepackage{subfig}
\def\BibTeX{{\rm B\kern-.05em{\sc i\kern-.025em b}\kern-.08em
    T\kern-.1667em\lower.7ex\hbox{E}\kern-.125emX}}
\begin{document}

%\title{Patience Is Profitable in Payment 
%Channel Networks}

\title{Understanding the Benefit of Being Patient in Payment Channel Networks}

%\title{Just Be More Patient}
\author{Qianlan Bai$^{\dag}$, Yuedong Xu$^+$,  Xin Wang$^{\dag}$\\

	{$^+$ School of Information Science and Engineering, Fudan University}\\
	
	{{$^{\dag}$ School of Computer Science and Technology, Fudan University}
	}
}

\maketitle

\begin{abstract}

Scaling blockchain efficiency is crucial to its widespread usage in which the payment channel is one of the most prominent approaches. With payment channels and the network they construct, two users can move some transactions off-chain in a predetermined duration to avoid expensive and time-consuming on-chain settlements. Existing work is devoted to designing high-throughput payment channel networks (PCNs) or efficient PCN routing policies to reduce the fee charged by intermediate nodes. In this paper, we investigate the PCN routing from a different perspective by answering whether the routing fee of transactions can be saved through being a bit more patient. The key idea is to reorder the processing sequence of atomic transactions, other than to handle each of them separately and immediately. We present two mechanisms, one is periodic transaction processing assisted by a PCN broker and the other is purely strategic waiting. In the former, all the incoming transactions in a short time interval are processed collectively. We formulate an optimization model to minimize their total routing fee and derive the optimal permutation of processing transactions as well as the routing policy for each of them. A \textit{Shapley value} based scheme is presented to redistribute the benefit of reordering among the transactions efficiently and fairly. In the latter, we model the waiting time of a strategic transaction on a single payment channel as the \textit{first passage time} problem in queueing theory when the transaction value is higher than the edge capacity upon its arrival. By capturing the capacity dynamics, we are able to calculate the recursive expression of waiting time distribution that is useful to gauge a user's cost of patience. 
Experimental results manifest that our cost redistribution mechanism can effectively save routing fees for all the transactions, and the waiting time distribution coincides with the model well.

% The application of Blockchain from decentralized to scale suffer from poor scalability. Payment channel is a effective solution to solve scalability problem and it allows nodes that do not have a direct channel to pay through multiple hops. Transactions can be safely conducted in PCN between multi-hop channels, depending on the HTLC mechanism. The intermediate channel can charge a certain fee as a reward to help with the transaction. Selecting the appropriate intermediate channels with minimum cost is an important problem in payment channel network. In this paper, we study how users can how users can pay less transaction fees by sacrificing part of their efficiency. First, 

% aa

% aa

% aa

% aa

% aa

% aa

% aa

% aa

% aa
\end{abstract}

%\begin{IEEEkeywords}
%component, formatting, style, styling, insert
%\end{IEEEkeywords}

\section{Introduction}

Since the advent of Bitcoin in 2008, we have witnessed the booming of various decentralized cryptocurrencies and the tremendous attentions they have gained. 
The historical transactions between cryptocurrency clients are recorded in a global and public data structure known as \emph{blockchain}. It is envisioned that blockchain technology together with digital payment will embrace many new fields such as manufacturing, international trading, healthcare, etc \cite{ref:application1}\cite{ref:application2}\cite{ref:application3}. Despite of their ever-going prosperity, cryptocurrencies suffer from poor scalability. For instance, Bitcoin \cite{ref:vbitcoin} processes 7 transactions per second and  Ethereum processes 15 transactions per second \cite{ref:vethereum}. In contrast, visa network \cite{ref:vvisa} can handle 1700 transactions per second, several orders of magnitude ahead of cryptocurrencies. The very low throughput hurdles its wide adoption. Scaling blockchain efficiency has become one of the most important issues that needs to be solved in order to ravel the transitional means of payment.

At present, there are several effective ways to improve the scalability of blockchain including payment channel \cite{ref:lightning network}, segregated witness\cite{ref:seg} and shard chain\cite{ref:sharding} in which the payment channel is the most prominent one. It is an off-chain solution that two cryptocurrency users are allowed to 
deposit tokens on the blockchain for the transactions between them in a predetermined duration. When establishing a payment channel, both parties agree on the amount of deposit, also called \textit{capacity}, which measures the maximum value of transfer from one to the other. In the course of a transaction, the sum of capacity in both directions remain the same but the unilateral capacities change. The users (or nodes interchangeably) that do not have a direct channel can chain multiple payment channels together into a payment channel network (PCN). 
Practical PCNs include Lightning network \cite{ref:lightning network} in the Bitcoin system, Raiden network \cite{ref:raiden network} in the Ethereum system and COMIT \cite{ref:comit} in cross-chain applications that utilize the Hash-time Lock Contract (HTLC) scheme. A transaction can be routed from the sender to the receiver via intermediate nodes as long as the channel capacities on this route are sufficient, or will be processed on the public chain otherwise. As a return, the intermediate nodes  will charge each transaction a certain amount of fee for the interconnections provided, and this amount is usually much lower than the overall cost  on the public chain. Selecting the appropriate intermediate channels with the minimum cost has become an important issue in a blockchain PCN system. 

The literature on route selection with the purpose of transaction cost reduction 
can be roughly categorized into two types. 
One is for an individual transaction.
Zhang et al. \cite{ref:cheapay} designed the Cheapay algorithm to find the cheapest available path with time and capacity constraints. A series of closely related works by Piatkivskyi et al. \cite{ref:split}, Rohrer et al. \cite{ref:split2} and Piatkivskyi et al. \cite{ref:split3} presented new mechanisms which divided a transaction and transmitted on different paths to resist the capacity constraints. 
Ren et al.\cite{ref:alpha fee} proposed a new charging mechanism to maintain the balance of capacity, where the fee rate is defined as a reciprocal of the exponential function of the current capacity,  so that more transactions can be processed successfully.
The other is to design the mechanism for a sequence of transactions.
 Wang et al. in \cite{ref:flash} proposed the Flash algorithm. By dividing the transaction into two categories according to the transaction amount, the large transaction mainly focuses on the transaction fee cost, while the small transaction mainly focuses on the detection path cost. 
 %The two kinds of transactions are processed
 %differently in order to reduce the total %cost of the transaction. 
 Varma and Maguluri utilized bilateral queues achieving as many transactions as possible can be processed successfully by using the concept of re-balancing on the chain\cite{ref:throught optimal}. Sivaraman et al. proposed the Spider network with the idea of packetizing transactions and adopting a multi-path transport protocol for high-throughput PCN routing \cite{ref:NSDI}. 
However, the literature puts the emphasis on either the minimum cost or the high efficiency, while little attention is paid to their delicate tradeoff. 

In this paper, we study the payment channel routing from a novel perspective: \emph{instead of pursuing the extreme efficiency, can the users save their transaction routing fees by being a bit more patient?} Existing PCNs process the incoming transactions atomically and instantaneously \cite{ref:NSDI}. Multiple transactions that arrive at different instants are routed on a first-come-first-serve (FCFS) basis. Two adverse effects might occur. 
One is that a transaction fails in the PCN and has to resort to the costly public chain if there is not a path with sufficient funds, the other is that the overall routing fee of a set of transactions (from/to different nodes) is usually high. As the underlying reason, PCNs are \textit{time evolving} with dynamic edge capacities. For instance, Alice and Bob has built a payment channel. After a transaction from Alice to Bob, the capacity of Alice to Bob decreases and that of Bob to Alice increases, and their total capacity remains unchanged. Therefore, the FCFS processing is myopic in that a transaction possible misses the cheap paths created by the subsequent transactions, or intercepts their cheap paths unfortunately. Combined with the atomic transfer of transaction values in full, the FCFS means further increases the total routing fee. Inspired by this observation, we propose to reorder the processing of incoming transactions to reduce the routing fees, other than to process them immediately.

Our first mechanism stipulates (by a PCN broker) that the incoming transactions are handled together at a fixed duration periodically. We formulate an optimization problem to minimize their total atomic routing fee whose output is the permutation of orders of transactions and the corresponding routing policy for each of them. Some transactions gain more, some gain less, while some others might lose. Under certain rules, we present a coalitional game framework to incentivize the form of a grand coalition using the famous Shapley value as the cost re-distribution mechanism. This mechanism is shown to be individual rational, efficient and fair such that all the transactions in the coalition benefit from the reduction of total routing fee, and the higher success rate of PCN processing is achieved.  

Our second mechanism is rather intuitive. A ``patient'' transaction can wait for the increase of the edge capacity if the initial capacity upon its arrival is below its transaction value. This simple approach is practical in the absence of the intervention by a PCN broker. However, computing how much time it should wait is a very challenging task even on a single PCN channel with bidirectional edges. We model the waiting time as the \textit{first passage time} in queueing theory, that is, the first time that the initial capacity increases above the transaction value, given the stochastic transaction arrival processes on both directions of a single payment channel. A novel stochastic model is built to capture the dynamics of edge capacities. We compute the recursive expression of the distribution of waiting time that is useful for the transaction sender to gauge his cost of patience.  Simulation results validate the accuracy of our model. 

Our major contributions are briefly summarized as below. 

\begin{itemize}
    \item We propose a novel idea of reordering the transactions actively or opportunistically, other than the chronological processing in order to reduce the PCN routing cost. 
    \item We present a periodic transaction processing scheme that yields an optimal transaction order, and formulate a coalitional game with Shapley value as the benefit redistribution mechanism. 
    
    \item We present a novel transient queueing model to capture the capacity dynamics of a payment channel, and calculate the recursive expression on the waiting time distribution of a strategic transaction. 
    %whose value is below the edge capacity upon its 
    %arrival.  
\end{itemize}

\section{Problem Description}
\label{sec:problem}
In this section, we describe the mathematical model of payment channel networks (PCNs) and the motivation of being patient in PCNs by blockchain transactions. 

\subsection{Temporal Network Model} 
A payment channel network can be represented as a time-dependent directed graph ${\mathcal{G}_t} = (\mathcal{N}, \mathcal{E},\mathcal{C,W},\mathcal{B})$, where $\mathcal{N}$ is the set of nodes and $\mathcal{E}$ is the set of edges. Each node $n_i\in \mathcal{N}$ represents a cryptocurrency account that has built one or more payment channel agreements with other nodes; each directed edge $e_{ij} \in \mathcal{E}$ represents a payment channel from node $n_i$ to $n_j$. The edge $e_{ij}$ is associated with a 3-tuple $(c_{ij}(t), w_{ij},b_{ij})$, where the first is the maximum amount of cryptocurrency coins $n_i$ can pay to $n_j$ at time $t$ and the second is the price per-unit of coin transfer charged by $n_j$ on this edge if it is a relay node. Let $b_{ij}$ be the (flat-rate) base fee of using the channel $e_{ij}$ as the relay, regardless the amount of transaction size. We denote by $X_k$ the $k^{th}$ transaction in the payment channel network that is expressed as a 4-tuple $X_k = (n_s(k), n_r(k), v_k, t_k)$. 
The sender and the receiver of $X_k$ are denoted by $n_s(k)$ and $n_r(k)$, respectively. Here, $v_k$ is the payment value measured by the number of coins and $t_k$ is the time instant that the transaction takes place. One primary difference between the PCN and the traditional communication network is that a payment channel between $n_i$ and $n_j$ contains two bidirectional edges whose sum of capacities is a constant. When a transaction from $n_i$ to $n_j$ has been processed successfully, the capacity of edge $e_{ij}$ decreases while that of edge $e_{ji}$ increases. For any pair of nodes $n_s$ and $n_r$ in the transaction $X_k$, a \textit{feasible path} is an end-to-end path on $\mathcal{G}_t$ whose edge capacities are sufficient to transfer the transaction value.

The business model of using payment channels is the following. If the nodes $n_s$ and $n_r$ form a payment channel directly, there is no need for $n_s$ to pay for using this channel when $n_r$ is the receiver. If the transaction $X_k$ is forwarded along a path $\mathcal{P}_k(t)$ that starts at $n_s$ and ends at $n_r$, the intermediate nodes should be paid for the coin transfer. For instance, consider the path $\mathcal{P}_k(t)=\{n_1,n_2,n_3,n_4\}$. The node $n_1$ will pay a certain amount fees to $n_2$ of using the channel $e_{23}$, and $n_2$ needs to pay to $n_3$ for using the channel $e_{34}$, but $n_3$ does not pay to the receiver $n_4$. {Hence, the total routing fee that $X_k$ need to pay can be expressed as $f=(w_{34}\cdot v_k+b_{34})+[ w_{23}\cdot (v_k+v_kw_{34}+b_{34})+b_{23}]$}\cite{ref:cheapay}\cite{ref:aodv}\cite{ref:tx fee1}. Formally, the total routing fee on the $\mathcal{P}_k$ at graph $\mathcal{G}_t$ is given by:
\begin{align}
\label{eq:Sfee}
f(X_k,\mathcal{G}_t)=
\begin{cases}
& \xi \quad \quad\quad\quad\text{$\mathcal{P}_k(t) = \varnothing$;}\\
&\sum\limits_{\substack{e_{ij}\in \mathcal{P}_k\\\forall n_i\neq n_s(k)}}(w_{ij}\cdot  V_k(i,j)+{b_{ij}})\;\;\text{otherwise.}
\end{cases}
\end{align}

Here, {$V_k(i,j)$ represent the actual amount that $X_k$ need to transfer through $e_{ij}$, including $v_k$ and the routing fees to be paid for subsequent edges.} When the value of a transaction exceeds the edge capacity, it can be sent to the public chain \cite{ref:NSDI} that will incur a fixed amount of payment plus a much longer confirmation time. Without loss of generality, we deem the processing cost of a transaction as a constant $\xi$ in the public chain that is higher than the routing fee in the PCN\cite{ref:public fee}. 
%{\color{red} $\xi$ \cite{ref:public fee} is 
%much more than the routing fee in PCN, that is
%because it includes not only the transaction 
%fee on public chain but the delay cost due to 
%confirming time.}

In the hop-by-hop value transfer,  the transaction fee need to transmit with the original transaction, which means $f(X_k,\mathcal{G}_t)+v_k$ is the actual value which $n_s(k)$ need to pay at time $t$.
For the intermediate edges on $\mathcal{P}_k(t)$, the capacity must be not less than the value and all the total charge for remaining routes. 
% Apparently, there may be more than one payment path that can fulfill a given transaction request. Therefore, it is necessary for user to minimize the transaction fee, while guaranteeing the \emph{public fee constrain} and \emph{capacity constrain}. In order to achieve this goal, 
the feasible flow paths should satisfy:
\begin{align}
\label{eq:single community}
    % &min \qquad  f(Tx_k,G_t)\\
    \begin{cases}
    %  &V_k+\sum\limits_{e_j^m \in \mathcal{P}_k^t\setminus \widetilde{P}(j)}(w_j^m \cdot V_k+{B_i^j} )\leq c_j^i, \forall e_j^i \in \mathcal{P}_k^t\\
    &V_k(i,j)\leq c_{ij}(t), \forall e_{ij} \in \mathcal{P}_k(t)\\
    &f(X_k,\mathcal{G}_t)\leq \xi
    \end{cases}
\end{align}

% where $\widetilde{P}(j)$ is the set which contains edges that is before $e_j^m$ in $\mathcal{P}_k^t$. For example, if $\mathcal{P}_k^t=n_0 \to n_1 \to n_2 \to \ldots \to n_l$, then for $n_1$, $\widetilde{P}(n_1)=n_2 \to \ldots \to n_l$.

% Algorithm 1 is designed to solve this optimal problem, it is based on Dijkstra algorithm. On the basis of the basic Dijkstra algorithm, the choice of the cheapest path in the next step increases the judgment of the path capacity. If it does not meet the capacity constrain, then even if the path can reduce the total cost, it cannot be used. Except for finding the cheapest path, this algorithm also update the topology of PCN, including the edges, weight and capacity.

% In Algorithm \ref{alg:Framwork}, line 1 is used to delete the useless path whose capacity is less than the value we want to transmit.
% $xxx$ shows PCN is spare network, deleting useless edges can decrease the running time, we assume the number of edges in $G$ is $m$ and the complexity is $\mathcal{O}(m^2)$.
% line 15-19 is used to update the topology of PCN. For the selected path, the capacity needs to be updated, which is shown as the decrease of the capacity of the path and the increase of the capacity of the reverse edge. If there is no reverse edge, a new reverse edge will be generated. 
% \end{definition}

We make a few assumptions to simplify our modeling efforts while complying with the real-world blockchain PCNs. They are commonly used in the literature. 

\begin{assumption}
\textit{(Information availability)} 
The global information including the network topology, the channel capacities and edge prices, is foreknown to all the nodes  \cite{ref:assumption11}\cite{ref:assumption12}\cite{ref:assumption13}.
\end{assumption}

\begin{assumption}
\textit{(Immediate Processing)} The transaction over a payment channel network takes effect immediately if it is scheduled for processing  \cite{ref:flash}\cite{ref:assumption21}. 
\end{assumption}

\begin{assumption}
\textit{(Indivisible Transaction)} A transaction cannot be splitted into multiple transactions of smaller values. 
\end{assumption}

The splitting of transactions may involve the multi-path routing problem that is usually more complicated. We only consider the indivisible transactions in this work, yet our problem and methodology are applicable to the divisible transactions. 

\subsection{Transaction Reordering} 

The original purpose of constructing payment channel networks is to speed up the processing of Bitcoin transactions. The 
state-of-the-art efforts are devoted to designing high-throughput payment channel systems without sacrificing cryptocurrency security \cite{ref:lightning network}. As a basic incentive to maintain the PCN, the payment channels may charge the sender of the transaction a small amount of \textit{routing fee}. As a consequence, the user is inclined to selecting a feasible path that yields the minimum total routing fee \cite{ref:cheapay}. 

Given a train of transactions sorted by their arrival times, 
the graph $\mathcal{G}_t$ changes after each successful transfer. In this sequential processing order, choosing the minimum cost path for each transaction might be myopic because the graph is dynamic. A transaction may have a chance to find another path with even lower cost if it can ``wait'' for some time. We hereby illustrate that being patient is ``egoistic'' or ``altruistic''. 

\begin{figure}[!ht]
		\vspace{-0.16in}
		\centering
		\includegraphics[width=0.35\textwidth]{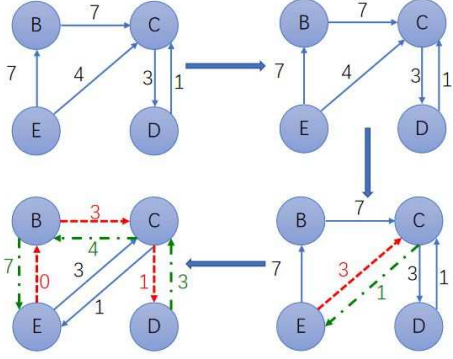}
		\caption{The process of $X_0=(D,C,3,0),X_1=(E,C,1,1),X_2=(E,D,2,2)$, ${b_{ij}}=1$, $w_{ij}=0.5  \quad        \forall e_{ij}\in \mathcal{E}$.}
		\label{fig:example1}
% 		\vspace{-0.2cm}
	\end{figure}

\begin{figure}[!ht]
		\vspace{-0.18in}
		\centering
		\includegraphics[width=0.35\textwidth]{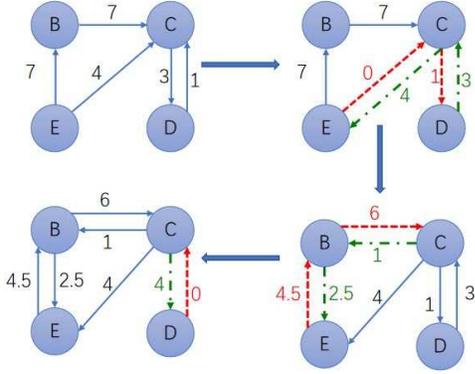}
		\caption{The process of $X_2=(E,D,2,2),X_1=(E,C,1,1),X_0=(D,C,3,0)$, ${b_{ij}}=1$, $w_{ij}=0.5 \quad \forall e_{ij}\in \mathcal{E}$.}
		\label{fig:example2}
		\vspace{-0.4cm}
	\end{figure}

{
% We use an example to explain the principle of the mechanism.
Fig.\ref{fig:example1} and Fig.\ref{fig:example2} shows the different routing fee that need to be paid for the same three transactions in different order. The three transactions are $X_0=(D,C,3,0),X_1=(E,C,1,1),X_2=(E,D,2,2)$. Without losing generality, base transaction fee is set as $1$ and each edge charge 50\% transaction amount. This setting has no influence for the problem which we want to illustrate. The red path (uniform dotted lines) represents the cheapest path through, the green path (uneven dotted line) represents the edges that change in the opposite direction. The number on the line indicates the capacity of the channel at the current time. Fig.\ref{fig:example1} shows the change in PCN topology when each transaction chooses to be processed as soon as it comes into being. For each transaction, the routing fee is $f(X_0,G_0)=\xi, f(X_1,G_1)=0, f(X_2,G_2)=5$. If $X_1$ chooses to be processed at time $\eta_1>2$ and $X_0$ chooses to be processed at time $\eta_0>\eta_1$, the sequential processing order is $X_2,X_1,X_0$, the change in PCN topology is shown in Fig.\ref{fig:example2}. In this order, routing fee is $f(X_0,G_{\eta_0})=0,f(X_1,G_{\eta_1})=1.5,f(X_2,G_2)=2$.
}

\noindent\textbf{Egoistic Waiting.}
{Egoistic waiting means that waiting can reduce the user's own routing cost. if $X_0$ chooses not to wait, it must be processed in public chain and the cost is $\xi$. Through waiting, $X_0$ can be processed after $X_1$ and $X_2$. there is an available path due to the successful processing of $X_1$ and $X_2$.The routing cost decrease for $X_0$ through waiting.}

\noindent\textbf{Altruistic Waiting.}
{Altruistic waiting is defined as the waiting of one user will cause its own routing cost to increase or remain unchanged, but other users' routing cost may decrease, leading to a decrease in the total cost.
If $X_1$ is processed at time $1$,  the capacity of $E \to C$ is sufficient and the routing fee is $0$. After waiting, $X_1$ can only choose $E\to B \to C$ due to the successful processing of $X_2$, the routing fee of $X_1$ increase to $1.5$. However, the routing fee of $X_2$ decrease to $2$ because the waiting of $X_1$. The total cost of $X_1$ and $X_2$ decrease. }

By reordering the transactions (other than first-come-first-serve (FCFS)), either some of the transactions can lower down their routing fees charged by the PCN, or these transactions as a whole can save a certain amount of routing fees. 
In order to take this advantage, there needs an 
incentive mechanism to encourage the users to be more patient. We hereby consider two scenarios, in which the former requires the intervention of the broker of the payment channel network, and the latter is completely compatible to the existing payment channel networks. 

\subsubsection{Periodic Transaction Processing (PTP)}
The broker of the PCN processes all the incoming transactions cyclically after \textbf{a fixed duration $\textbf{T}$}. An optimal reordering policy will be implemented to minimize the total transaction cost, and the total routing fee will be redistributed among these transactions. As the underlying principle, no transaction will receive a higher routing fee in the reordered processing than in the FCFS  processing. 

\subsubsection{Strategic Patience (SP)}
When a myopic user finds that the cheapest payment channel does not have the sufficient capacity to transfer his transaction value, he may resort to an expensive channel or the public chain. While a strategic user can predict how much time he can wait until the capacity of the cheap channel is greater than his transaction value, given the arrival patterns of the transactions at both directions of the channel.

\vspace{-0.1cm}
\begin{algorithm}[!htp]
  \caption{ Framework of calculating the cheapest path.}  
  \label{alg:Framwork}  
  \textbf{Input:} Transaction, $X=(n_s,n_r,v)$;
        Topology of PCN, $G_0$;\\
        \textbf{Output:}      Transaction fee, $f$;\,\,\, Path, $\mathcal{P}$; New topology, $G$;\\
  \begin{algorithmic}[1]
\STATE $G_1\gets G_0$ for all edges with $(c_{ij}<v)$ removed;
\STATE $N=G_1.nodes$; $E=G_1.edges$;
    \STATE  $R\gets \{n_r\}$; $S\gets{N\setminus n_r}$; 
    \STATE Initialize $result$ as $\{(fee_i,path_i)\}$,\\ for every $e_{ij} \in E$, $fee_i\gets w_{ij}\cdot v+b_{ij}, path_i\gets [n_j,n_i]$;\\ otherwise, $fee_i \gets \xi, path_i\gets [n_i]$. 
      \STATE \textbf{${w_{{n_s}j}\gets 0}$, $b_{{n_s}j}\gets 0$, for each $n_i \in n_s.neighbor$}; 
    % $r_i^{n_r}=c_i^{n_r}$;\\
    %   \ENDIF\\
      \WHILE{$result$ was updated and $S$ is not empty}
      \STATE $n_i\gets \underset{n_k \in R}{\arg\min}{fee_k}$
    %  \STATE $n_i\gets \{n_i\in R| {min(result[n_i].len)}\}$ 
       \FOR{$n_j$ in $n_i.pre$} 
    %   \FOR{ all $n_i$ in $result$}
    %  \FOR{all $n_j$ in $G[n_i]$}
      \IF{$fee_i+ w_{ji}\cdot(v+fee_i)+b_{ji}<fee_j$ \textbf{and} $v+w_{ji}\cdot (v+fee_i)\leq c_{ji}$}
      \STATE $fee_j\gets fee_i+w_{ji}\cdot(v+fee_i)+b_{ji}$;
      \STATE $path_j\gets path_i.append(n_j) $;
        \STATE $R.append(n_j); S.remove(n_j)$;
      \ENDIF
      \ENDFOR
    %   \ENDFOR
      \ENDWHILE 

$fee\gets fee_s$;
$path\gets path_s$;\\
% \textit{\%update the capacity of $G_0$};\\
% index=-1; inter=0; \textit{\% index of nodes in the path}

% \WHILE{index!=$-|\mathcal{P}|$}
\FOR{$e_{ij}\in path$}
% \STATE $i\gets path[index-1],j\gets path[index]$;
\STATE $c_{ij} \gets c_{ij}-(v+fee_i)$;\\
\STATE $c_{ji}\gets c_{ji}+(v+fee_i)$;\\
% $inter \gets w_{ij}\cdot(v+inter+b_{ij})$;
% $index \gets index-1$;
\ENDFOR\\
    \RETURN  $f\gets fee_s,\mathcal{P}\gets path_s$, $G\gets G_0$ ;
     \end{algorithmic}  
\end{algorithm}

% \subsection{Cooperation mechanism design}

%  \textbf{Therefore, we can find patience can reduce not only their own transaction fees, but also those of other transactions, and even some transactions that cannot be processed immediately on PCN can be processed successfully.}

%  We hope to design a mechanism that can sacrifice the processing speed of some transactions to achieve less transaction fee. The mechanism is that each transaction is not processed at the moment the transaction arrives, but is processed every $T$ by the system.
%  At the moment that transaction generate, they can choose join in coalition or not.
% If joining in coalition, system will reorder the transactions to pay for less transaction fee. 

\section{Coalition Mechanism design}
\label{sec:shapley}
In this section, we formulate a cooperative game and formally define the benefit distribution mechanism that is compatible to the optimal transaction reordering principle.

\subsection{Cooperative game in PCN} 
Cooperative game (or coalition game narrowly) is a mathematical theory in revealing the behaviors of rational players in a cooperative setting. 
The players make agreements among themselves to form coalitions that affect their strategies and utilities, as opposed to the non-cooperative games. 
In what follows, we formulate the benefit (or cost) redistribution mechanism as a cooperative game, namely $CG$. 
\begin{itemize}
    \item \textbf{Player:} A user who initiates a transaction is regarded as a player. If not mentioned explicitly, the transaction is equivalent to the user so that the set of players are expressed as $\mathbf{X}:=\{X_0,X_1,\cdots,X_{N-1}\}$ with the arrival times $t_n \leq T$, $\forall \; 0\leq n \leq N-1$. 
    
    \item \textbf{Worth function:} A function $\psi(S)$ where $S$ is an arbitrary non-empty subset of $\mathbf{X}$. 
    
\end{itemize}
We denote $\psi(S)$ as the \emph{worth function}, which measures the benefit produced by coalitions. The redistribution mechanism in the coalition is Shapley value function.

\begin{definition}[Grand coalition]
If every player chooses to join in coalition, the coalition is called as grand coalition.
\end{definition}

\subsection{Worth of $S$}
The calculation of the worth function $\psi(S)$ is special in our PCN coalitional game. 
First of all, $\psi(S)$ is an outcome of minimum cost routing problem that demands an appropriate algorithm to generate this value. Second, the coalition is expressed in the form of reordering the transactions so that the sequence number of processing this coalition as a whole needs to be decided. Third, there might exist the ``free-riding'' phenomenon in the game. To facilitate the form of physically meaningful coalition(s), a set of rules are enforced.

\textbf{Rule 1:} The sequence order of a coalition $S \in \mathbf{X}$ is determined by that of the earliest transaction in $S$. Denote by $\eta_S$ the order of coalition $S$, and by $\eta_k$ the order of $X_k$ in the original set of transactions. There are 
\begin{align}
\label{eq:priority}
&\eta_k=\arg\min_j\{t_k | X_k \notin S\};\\
\label{eq:priorityS}
&\eta_S=\arg\min_j\{t_j | X_j \in S\}.
\end{align}
For instance, given four transactions $X_0,X_1,X_2,X_3$, where $X_1$ and $X_3$ form a coalition $S$, $\eta_0=0, \eta_S=1, \eta_3=3$. In other word,  the processing sequence is $X_0$, $S$, $X_2$ accordingly. This is to say, the coalition does not affect the processing of the transactions who arrive earlier than all the coalitional players. 

\textbf{Rule 2:}
In $S$, the sequence of players should be rearranged so as to minimize the total routing fee. 
Consider a grand coalition $S=\{X_0, X_1, \cdots, X_{K-1}\}$, the PCN graph is redefined as $\mathcal{G}_k$ before processing the transaction $X_k$. The optimal routing for a particular $X_k$ in the current topology of PCN is expressed below. 
\begin{align}
\label{eq:optimal function}
   \min \quad\quad   &\sum\limits_{(i,j):e_{ij}\in \mathcal{E}}f_{ij}(X_k)\gamma_{ij}(X_k)
   \end{align}
    %   & s.t.\sum\limits_{X_k\in S}V_k(i,j)\cdot e_{ij}(k) -\sum\limits_{X_q \in S}V_q(j,i)\cdot e_{ji}(q)\leqslant c_{ij}, \forall e_{ij}\in E.
    %   &\eta_S\leq min(t_i+\zeta_i), \forall Tx_i\in Tx.
\begin{align}
\label{eq:gamma}
    s.t.\quad     \gamma_{ij}(X_k):=
    \begin{cases}
 &1,\quad \text{if $X_k$ is routed along $e_{ij}$};\\
 &0,\quad \text{else}.\\
    \end{cases} 
\end{align}
\begin{align}
\label{eq:flow balance}
          &\sum\limits_{j\in \mathcal{N}}\gamma_{ij}(X_k)-\sum\limits_{j\in\mathcal{N}}\gamma_{ji}(X_k)=:
        \begin{cases}
         &1,\quad n_i = n_s(k);\\
         &-1,\quad n_i =n_r(k);\\
         &0,\quad \text{else}.
        \end{cases} 
\end{align}
\begin{align}
\label{eq:capacity constrain}
    &\gamma_{ij}(X_k)(v_k+f_{ij}(X_k))\leq c_{ij}(\eta_k),\quad \forall e_{ij}\in \mathcal{E};\\
    \label{eq:fee1}
       & f_{ij}(X_k)=\gamma_{ij}(X_k)[\sum\limits_{h\in \mathcal{N}}\gamma_{jh}(X_k)(w_{jh}(v_k+f_{jh}(X_k))+b_{jh})]\\
       \label{eq:fee2}
       &f_{ij}(X_k)=0, \quad \text{if $n_j = n_r(k)$};\\
    &\sum\limits_{(i,j):e_{ij}\in \mathcal{E}}f_{ij}(X_k)\gamma_{ij}(X_k)\leq \xi.
\end{align}
The objective is the sum of the routing fees on all the edges. The binary variable $\gamma_{ij}(X_k)$ in Eq.\eqref{eq:gamma} indicates the usage of an edge, Eq.\eqref{eq:flow balance} represents the flow balance conditions, and Eq.\eqref{eq:capacity constrain} presents the maximum amount of values that can be processed. 
Eq.\eqref{eq:fee1}$\sim$\eqref{eq:fee2} refers to the transaction fee calculations, i.e. $f_{ij}(X_k)$ represents the routing fee that $X_k$ need to pay to other nodes after the edge $e_{ij}$. 
Actually, the main constraints are about the selection of feasible paths, and we need to find the cheapest one among the set of feasible paths. If such a path leads to a higher routing cost than $\xi$, the public chain will be selected for the value transfer. 

Algorithm 1 is used to calculate the cheapest path for a single transaction and to update the PCN topology(Eq\ref{eq:optimal function}).  Algorithm 2 is designed to calculate the minimum routing fee that coalition need to pay.
%  Algorithm 1 is used to calculate the cheapest path for single transaction and update the topology of PCN.
 According to Algorithm 2, we can find that we should ensure as much as possible transactions can be processed successfully firstly. Under this premise, the value that can be transferred should be as much as possible.

We can get
\begin{align}
    \psi(S)= \sum\limits_{X_k\in S}f(X_k,G_{t_k})-f(X_k,G_{\eta_S})
\end{align}
% \begin{equation}
% x_k=
% \begin{cases}
% 0& \text{if $Tx_k$ is processed successfully}\\
% 1& \text{otherwise}
% \end{cases}
% \end{equation}

% Obviously, this problem is an \emph{optimal problem}.
% It can be calculated by adjusting the transaction order and calling algorithm 1 in turn. This optimal function ensure the maximum number of transactions which are processed successfully. 
There are some properties about $\psi(S)$.
\begin{itemize}
       \item \emph{Cohesive:}{ A coalitional game $<\mathbf{X}, \psi>$ with transferable payoff is cohesive if }
\begin{displaymath}
      \psi(\mathbf{X}) \geq \sum_{k=1}^{n} \psi(S_k)
\end{displaymath}
 \emph{for every partition $\{S_1, \ldots, S_n\}$ of $\mathbf{X}$.}
    \item \emph{Weak superadditivity:} In a cooperation game $<\mathbf{X}, \psi>$, for any $S$, $S_1$, $S \subset \mathbf{X}, S_1 \subset S$, then there must be $\psi(S_1)\leq \psi(S)$, it is said that the characteristic function $\psi$ satisfies {weak superadditivity}.
\end{itemize}

The total benefit of coalition will increase because of the existence of \emph{altruistic waiting} user. It is necessary for \emph{egoistic waiting} user share part of their benefit with altruistic waiting user to compensate them for their losses.
But there are some egoistic waiting users who want to monopolize the benefit and not join the coalition.
This kind of players is called as \emph{Free-rider}. 
In the social sciences, the free-rider problem is a type of market failure that occurs when those who benefit from resources, public goods (such as public roads or hospitals), or services of a communal nature do not pay for them[1] or under-pay. Free riders are a problem because while not paying for the good (either directly through fees or tolls or indirectly through taxes), they may continue to access or use it. We use an example to illustrate this problem and establish a constrain to avoid this kind of phenomenon. 
 	\begin{figure}[!ht]
		\vspace{-0.3cm}
		\centering
		\includegraphics[width=0.4\textwidth]{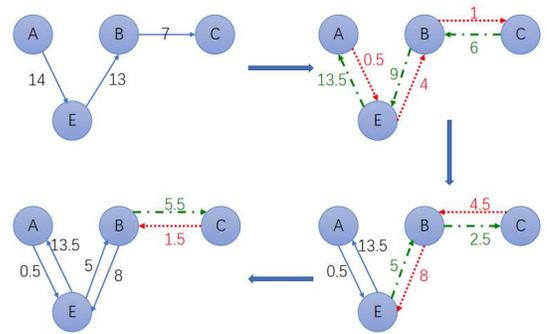}
		\caption{The process of $X_2=(A,C,6,2),X_0=(C,E,1,0),X_1=(C,B,3,1)$, ${b_{ij}}=0$, $w_{ij}=0.5 \quad \forall e_{ij}\in {E}$.}
		\label{fig:free-rider}
		\vspace{-0.3cm}
	\end{figure}

As Fig.\ref{fig:free-rider} shows, there are three transactions, $X_0=(C,E,1,0),X_1=(C,B,3,1),X_2=(A,C,6,2)$, to simplify, we set ${b_{ij}}=0$ and all of the $w_{ij}=0.5$. If each transaction choose not to cooperate, the benefit of three transactions are $0,0,0$ separately. If $X_2$ and $X_0$ choose to cooperate, according the priority mechanism, they should be processed firstly and the processing order is $X_2, X_0,X_1$. The benefit are $\xi-0.5, \xi, 0$ for $X_0,X_1$ and $X_2$. $X_0$ is egoistic waiting user and according the common sense, he should share some benefit with $X_2$.  $X_1$ also benefit but he does not share his earning with $X_2$ because of not joining in coalition. We call $X_1$ \emph{Free-rider}. $X_2$ is the altruistic waiting user and he is not paid for his service to free-rider. Free-rider will lead to the instability of the coalition, so we need to set up an effective mechanism to resist this phenomenon.  

% \flushleft
% \newtheorem{definition}{Definition}[section]
\begin{definition}[Free-rider solution]
{The topology of PCN is stored. The path of the player who does not join in $S$ is checked and the cheaper path which is created by $S$ is prohibited from using them.} 
\end{definition}

Free-rider solution can protect the payoff of $S$ and to avoid loss of income, middle nodes will choose obey it.

% \emph{Remark:} The property $f(Tx_k,G_t)<<\xi$ can ensure that system 
% will not give up some small transactions in order to obtain higher interest, so as to ensure the fairness of the players.

\begin{algorithm}[!htp]  
  \caption{ Calculating the minimum routing fee of coalition.}  
  \label{alg:optimal function}  
  \textbf{Input:}  Coalition, $S=\{X_0,X_1,\ldots, X_n\}$; \, Topology of PCN, $G_{0}$;\\
\textbf{Output:}$\{(fee_i,path_i)\}$ for all $X_i \in S$;\\
  \begin{algorithmic}[1]
  \STATE $ X_{list}=\{Req_i|Req_i \in Permutations(S)\}$;
\STATE $G\gets G_{0}$;
\FOR{all $Req_i$}
 \FOR{all $X_j \in Req_i$}
   \STATE $f,\mathcal{P},G=\textbf{Algorithm 1}(X_j,G)$;
      \STATE $result_{all}$.append($f,\mathcal{P}$);
 \ENDFOR
\ENDFOR
\STATE $result\gets$ $r_i\in result_{all}$ with largest number of successful transactions. 
   \IF{Number of successful transactions is same}
   \STATE $result\gets$ $r_i\in result_{all}$ with more successful transfer amount.
   \ENDIF
    \RETURN  $result$;
     \end{algorithmic}  
\end{algorithm}

\subsection{Benefit Redistribution via Shapley Value}

The reordering of the players may cause some of them pay less routing fee and while some others pay more. In order to incentive the players to form a grand coalition, the redistribution of the benefit is inevitable. It needs to be \textit{fair} to all the players, and yields a \textit{unique} payoff vector known as the \textit{value} of the coalitional game\cite{ref:a course}. 

\begin{definition}
    A \textit{benefit redistribution mechanism} is an operator $\phi$ on a payment channel network $<\mathbf{X}, \psi>$ that allocates a cost vector $\phi(\mathbf{X}, \psi) = (\phi_0,\phi_1,\cdots, \phi_{N-1})$ in $\mathbb{R}^{N}$ for all the players.
\end{definition} 

We hereby design a benefit redistribution mechanism with the following desirable properties among the players.

% \newtheorem{property}{Property}%[section]
% \begin{property}
%  We call it as 
% \end{property}

\newtheorem{property}{Property}
\begin{property}[Rationality]
\emph{If $\phi_i(S)\geqslant \psi(\{X_i\})$, it is \textbf{individual rationality};
If $\sum\limits_{X_i\in S}(\phi_i(S))=\psi(S)$, it is \textbf{group rationality}}.

Individual rationality motivates the players to join the coalition and cooperate accordingly. Group rationality requires that the profit assigned equals the profit received from the coalition.  
\end{property}
% \flushleft
% \newtheorem{Lemma}{Lemma}[section]
% \begin{Lemma}
% \emph{If effectiveness cannot be realized, then individual rationality will not be realized.}
% \end{Lemma}

% \textbf{Proof:} According to effectiveness, for any coalition $S$ we can get 
% \begin{displaymath}
%       \psi(S) \geq \sum_{i}V(\{i\})=\sum_{i}f(Tx_i, G_{t_i}), \forall Tx_i \in S.
% \end{displaymath}
% If effectiveness can not be satisfied, then it means there is bound to be 
% \begin{displaymath}
%       \psi(S) \leq \sum_{i}f(Tx_i, G_{t_i}), \forall Tx_i \in S.
% \end{displaymath}
% According to group rationality, we can get
% \begin{displaymath}
%       \sum_{i}\phi_{Tx_i}(S) \leq \sum_{i}f(Tx_i, G_{t_i}), \forall Tx_i \in S.
% \end{displaymath}
% This means that there are bound to be some players whose profits from cooperation are even lower than those without cooperation. Therefore, the individual rationality is violated.

\begin{property}[Balanced contribution]
\emph{A value $\phi$ satisfies the  \textbf{balanced contributions property} if for every coalitional game with transferable benefit $<\mathbf{X}, \psi>$ we have 
\begin{displaymath}
     \phi_i(\mathbf{X})-\phi_i(\mathbf{X} \backslash \{X_j\})= 
  \phi_j(\mathbf{X})-\phi_j(\mathbf{X}\backslash\{X_i\})   
\end{displaymath}
where $X_i\in \mathbf{X}$ and $X_j \in \mathbf{X}$.
}
\end{property}

The property of balanced contributions addresses the fairness between any pair of transactions in $\mathbf{X}$. If we start with a set of two transactions $(\mathbf{X}, \psi)=(\{X_1,X_2\},\psi)$, the gain from cooperation is $\psi(\mathbf{X})-\psi({X_1})-\psi({X_2})$. Thus, the egalitarian solution is 
\begin{displaymath}
      \phi_i(\mathbf{X},\psi)=\psi(\{i\})+\frac{1}{2}\left[\psi(\mathbf{X})-\psi(\{X_1\})-\psi(\{X_2\})\right].
\end{displaymath}
% The unique value that satisfies the balanced contributions property is the Shapely value.

\begin{property}[Symmetry]
\emph{If $\psi(S \cup {X_i})=\psi(S \cup {X_j})$, for all $S \in \mathbf{X} \backslash \{X_i, X_j\}$, then $\phi_i(S)=\phi_j(S)$.}
\end{property}

The symmetry property requires that if two players contribute the same to every subset of other players, they should receive the same amount of cost.

\begin{property}[Dummy]\label{dum}
\emph{If $i$ is a dummy player in coalition $S$ then $\phi(X_k)= \psi(\{X_k\})$.}
\end{property}

In our game, if a player does not contribution to the reduction of routing fee, i.e. $\psi(S)+\psi(\{X_i\})=\psi(S \cup {X_i})$, the payoff of this player of joining the coalition is identical to that of not joining it. 

\begin{property}[Additivity]
\emph{For any two game ($S, \psi_1$) and ($S, \psi_2$) we have $\phi_i(\psi_1+\psi_2)=\phi_i(\psi_1)+\phi_i(\psi_2)$ for all $i \in S$, where $\psi_1+\psi_2$ is the game defined by $(\psi_1+\psi_2)(S)=\psi_1(S)+\psi_2(S)$ for every coalition $S$.}
\end{property}
%\begin{property}
%\emph{For any two coalitional games $<Tx, V>$ and $<Tx, W>$, we have %$\psi_i(V+W)=\psi_i(V)+\psi_i(W)$ for all $i\in Tx$, where $V+W$ is the game defined by %$(V+W)(S)=V(S)+W(S)$ for every coalition $S$. We call it as \textbf{additivity}.}
%\end{property}
Additivity ensures that even if charging rate of some edges changes, our redistribution mechanism is still available.

The \textbf{\emph{Shapley value}} is the unique value that satisfies all five properties. Then, it is defined as follows.\cite{ref:shapely value}
\begin{definition}[Marginal contribution]
{The \emph{marginal contribution} of player $i$ to any coalition $S$ with $i \not\in S$ in the game $<X, \psi>$ to be
\begin{displaymath}
   \bigtriangleup_i(S)=\psi(S\cup\{i\})-\psi(S)   
\end{displaymath}}
\end{definition}

\begin{definition}[Shapley value]
{The Shapley value $\phi$ is defined by the condition
\begin{align}
\label{eq:shapely value}
\phi_i(S)=\sum\limits_{S_1\subseteq S}\frac{(|S|-|S_1|)!(|S_1|-1)!}{|S|!}
   \bigtriangleup_i(S_1\cup \{X_i\}),\,
% [\psi(S_1)-\psi(S_1\backslash\{i\})]
\forall X_i\in S.
    %   \phi_i(X)=\frac{1}{\left|X\right|!}\sum_{R\in \mathcal{R}}(\bigtriangleup_i(S_i(R))),\quad \forall i\in X.
\end{align}
}
\end{definition}
where $|S|$ and $|S_1|$ represent the numbers of players in $S$ and $S_1$, respectively. Actually, Shapley value represents the expected marginal contribution over all orders of this player to the set of players who precede him.
% where $\mathcal{R}$ is the set of all $\left|X\right|!$ orderings of $X$ and $S_i(R)$ is the set of players preceding $i$ in the ordering $R$. 

% About \emph{symmetry, dummy}, the proof can be found in
% \begin{definition}
%  \emph{In a cooperation game $<Tx, \psi>$, for any coalition $S,T \subset Tx, S\bigcap T=\varnothing$, there will be
% $ \psi(S)+\psi(T)\leq \psi(S \bigcup T)$, it is said that the characteristic function $\psi$ satisfies \textbf{superadditivity}}.
% \end{definition}

% \begin{definition}
%  \emph{In a cooperation game $<Tx, \psi>$, for any coalition $S,T \subseteq Tx$, there will be
% $ \psi(S)+\psi(T)\leq \psi(S \bigcup T)+\psi(S \bigcap T)$, it is said that the characteristic function $\psi$ satisfies \textbf{convex}}.
% \end{definition}
% If $\psi$ satisfy $convex$, then it must be superadditivity.

  \flushleft
 \begin{lemma}
 \label{superindividual}
{ $\psi$ satisfy weak superadditivity and the corresponding Shapley value satisfies the individual rationality.} 
 \end{lemma}
 \textbf{Proof:} Here, $\psi$ satisfies weak superadditivity, then:
 \begin{displaymath}
       \psi(S\cup {i})\geq \psi(S)+\psi(\{i\}), \forall S\in X,  i \in X \backslash S 
 \end{displaymath}
 Therefore $\bigtriangleup_i(S) \geq \psi(\{i\})$. According Eq\eqref{eq:shapely value},
 we can get that $\phi_{i}(S)$ satisfies individual rationality. This lemma is proved.
 
%  \begin{lemma}
% \emph{In CG, $\phi_i(S)$ satisfy individual rationality.}\cite{ref:individual rationality}
% \end{lemma}
% \textbf{Proof:}
% According to Eq. \ref{eq:optimal function}, we can get
% \begin{eqnarray}
%          \psi(S) \geq 0, \forall S\subseteq X
% \end{eqnarray}
% Which means
% \begin{align}
% \psi(S)\geq \psi(S_1), \forall S_1 \subset S.
% \end{align}

% Therefore, $\psi(S)$ satisfy weakly superadditivity. According to Lemma \ref{superindividual}, we can prove this lemma.

% \subsection{Free-rider}

{

\subsection{Operation Procedure}
We have shown the cooperation mechanism and the redistribution function in the last three subsections. In this subsection, we address the operation procedure of \emph{PTP}. Let us briefly introduce the implementation process of this mechanism.
% As a distributed anonymous system, we need to consider how to distribute the revenue of the coalition honestly and effectively in the payment channel network. A smart contract is a self-executing contract with the terms of the agreement between buyer and seller being directly written into lines of code. Smart contracts permit trusted transactions and agreements to be carried out among disparate, anonymous parties without the need for a central authority, legal system, or external enforcement mechanism. Based on the above reasons, we choose the intelligent contract to realize the income distribution.
% The system framework is shown in Fig \ref{fig:smart contract}.
%  	\begin{figure}[!ht]
% 		%\vspace{-0.18in}
% 		\centering
% 		\includegraphics[width=3in]{smart contract.PNG}
% 		\caption{Framework of smart contract.}
% 		\label{fig:smart contract}
% 		%\vspace{-0.2in}
% 	\end{figure}

\begin{itemize}
    \item Step1: Before $T$, each transaction $X_i$ need to chooses joining in the coalition or not at $t_i$. No matter they join or not, they need to submit information to system, including their sender, receiver, value. The arrival time is automatically recorded by the system.
    \item Step2: At time $T$, system first calculate the routing fee for each transaction based on the first-come-first-serve. This result can get through the information they submit.
    \item Step3: Then, system calculate the priority of each transaction through Eq\eqref{eq:priorityS} and Eq\eqref{eq:priority}. 
    \item Step4: According to the result of priority, system begin to process the transactions. For coalition $S$, it will reorder the players to get the minimum routing fee they need to pay. Algorithm 2 is used to solve this problem. 
    \item Step5: System need to distribution the payoff to each player in $S$ using $\phi_i(S)$.
    \item Step6: System initialize the time to 0 until the next $T$ arrives, repeating from step1.
\end{itemize}
\emph{Remark:} All operations can be done through smart contracts, so it does not break the principle of decentralization. The system need to store topology of PCN at initial time to defend free-rider.
%  	\begin{figure}[!ht]
% 		%\vspace{-0.18in}
% 		\centering
% 		\includegraphics[width=3in]{single contract.JPG}
% 		\caption{Structure of smart contract.}
% 		\label{fig:single contract}
% 		%\vspace{-0.2in}
% 	\end{figure}
}

\section{Stochastic Waiting Time}
\label{sec:stochastic}

In this section, we formulate the stochastic model of capacity dynamics on a simplified PCN and calculate the waiting time distribution of a strategic user. 

\subsection{Stochastic Capacity Model}
\begin{figure}[!ht]
		\vspace{-0.4cm}
		\centering
		\includegraphics[width=0.5\textwidth]{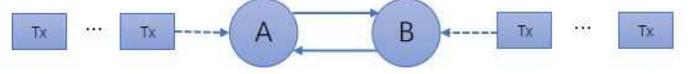}
		\caption{A simple network for analyzing stochastic waiting time.}
		\label{fig:single channel example}
		\vspace{-0.3cm}
	\end{figure}

When there is absence of the intervention of the PCN broker, the cost redistribution will be difficult to implement because the utility transfer among users cannot be enforced. Now we will show that an individual ``patient'' transaction can achieve a lower cost by waiting for the  feasibility of a payment channel when the capacity of this channel is below the transaction value in the very beginning. 
We consider a simplified PCN with only two nodes forming a payment channel in Fig.\ref{fig:single channel example} that can be easily generalized to the network with parallel payment channels. Node $A$ (resp. node $B$) is in charge of processing the left-side (resp. right-side) transactions on edge $e_{AB}$ (resp. edge $e_{BA}$). The successful transactions from $A$ to $B$ increase the capacity $c_{BA}$ and decrease the capacity $c_{AB}$, and vice versa. When a transaction finds the capacity $c_{AB}$ insufficient, it can keep patient until $c_{AB}$ is larger than its transaction value. An interesting question is how much time a transaction needs to wait before the successful processing, given the stochastic arrivals of other transactions on both sides of the channel.

\noindent\textbf{Capacity Dynamics.} The time axis is set to $t=0$ when a transaction arrives to node $A$ and will be transferred to node $B$ through edge $e_{AB}$. If the initial capacity $u$ is no less  than the transaction value $v$, it is processed immediately. Otherwise, it will wait for until $c_{AB}$ is greater than $v$. Without loss of generality, we denote this transaction as the tagged transaction $\hat{X}$. Denote by $U_t$ the capacity of edge $e_{AB}$ at time $t$, there exists 
\begin{align}
\label{eq:surplus}
    &U_t=u+\sum\limits_{i=0}^{N_2(t)}{v_{2i}}-\sum\limits_{i=0}^{N_1(t)}{v_{1i}}
\end{align}
where $v_{1i}$ indicates the value of the $i^{th}$ transaction from $A$ to $B$ and $v_{2i}$ indicates that of the $i^{th}$ transaction from $B$ to $A$. Here, 
$N_1(t)$ and $N_2(t)$ are the numbers of transactions on edge $e_{AB}$ and $e_{BA}$ by time $t$ respectively. We make the following assumptions. 
\begin{itemize}
    \item The arrival process of transactions on edge $e_{AB}$ is the Poisson process $\{N_1(t):t\geq0\}$ with parameter $\lambda_1$, and that on edge $e_{BA}$ is the Poisson process $\{N_2(t):t\geq0\}$ with parameter $\lambda_2$. 
    
    \item The transaction values on both directions, i.e. $v_{1i}$ and $v_{2i}$, are independent and identically distributed (i.i.d) with the probability density function $g(v)$.
\end{itemize} 
The assumption of Poisson arrival is commonly adopted in decentralized payment systems \cite{ref:poisson payment}, and the bilateral transactions are deemed to have the same distribution of values but with different arrival rates. 
Therefore, the evolution of $\{U_t; t\geq 0\}$ is a \textit{compound} Poisson process. 

The waiting time of the transaction $\hat{X}$ is actually the duration between $0$ and the instant that the capacity $c_{AB}$ is greater than $v$ for the first time. Thus, the waiting time can be modeled as the \textit{first passage time} of $U_t$ to $v$ in queueing theory. Formally, we denote $\hat{t}$ as the waiting time that has 
\begin{align}
\widehat{t}=\underset{t}\arg\min\{t\; |\;{u-\sum\limits_{i=0}^{N_1({t})}{v_{1i}}+\sum\limits_{i=0}^{N_2({t})}{v_{2i}}\geq v}\}
 \label{eq:third model}
        % & \alpha*f(Tx_k,G_{\widehat{t_k}})+\beta*d(Tx_k,\widehat{t_k}) < \xi
    %   \label{eq:single channel}
\end{align}
Given the stochastic arrival of transactions on the edges $e_{AB}$ and $e_{BA}$, $\hat{t}$ is a random variable by nature.

Before calculating the distribution of $\hat{t}$, we provide the precondition of waiting. The sum of $c_{AB}$ and $c_{BA}$ has been decided upon the construction of the payment channel so that the transaction $\widehat{X}$ cannot be processed in this channel if this sum is below $v$. We provides the following lemma with regard to the expected waiting time but omitting the proof due to its simplicity. 
\begin{lemma}
The expected waiting time $\mathbb{E}[\hat{t}|u]$ is calculated by
\begin{equation}
\mathbb{E}[\hat{t}] = 
\begin{cases}
\frac{v - u}{(\lambda_2 - \lambda_1)\mu}& \text{$\lambda_2 > \lambda_1$.}\\
\infty& \text{otherwise.}
\end{cases}
\end{equation}

\end{lemma}

%The approximated average waiting time can be 
%easily given by 

\subsection{Computing Waiting Time Distribution} 

The expected waiting time overlooks the stochastic behavior of transaction arrivals and the random transaction values, which is not sufficient to quantify the characteristic of the waiting time. We are more interested in how the chance of the successful processing increases over the waiting time. Hereby we analyze the probability distribution of the waiting time. Denote by $\Phi(v_k, u, t)$ the probability of a transaction being processed by time $t$
\begin{eqnarray}
\Phi(v_k, u, t) = \textbf{Pr}\{\hat{t} \leq t| U_0 = u, v = v_k\}.
\end{eqnarray}
where $u$ is the initial capacity observed by $X$. 

\begin{figure}[!ht]
		%\vspace{-0.18in}
		\centering
		\includegraphics[width=0.4\textwidth]{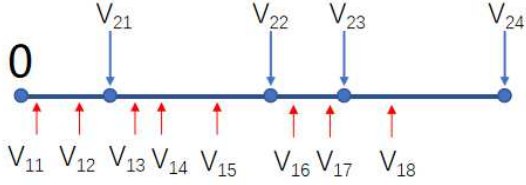}
		\caption{Example of arrival transactions on time-slot.}
		\label{fig:timeslot}
		\vspace{-0.5cm}
	\end{figure}
	
Computing $\Phi(v_k, u, t)$ is a very challenging task. We use a timeline in Fig.\ref{fig:timeslot} to illustrate the possible events that inspires our basic idea. One can observe that the first passage event must occur at the instant of processing a transaction on edge $e_{BA}$. Then, we can compute  $\varphi(r, n)$ which is the probability that the first passage event takes place at the arrival of the $n^{th}$ transaction on edge $e_{BA}$, where $r=v_k-u$. 
The calculation of $\Phi(v_k, u, t)$ is thus transformed into the union of $\varphi(r, n)$ for all $n$ mutually exclusive events that happen before time $t$. 
During the inter-arrival time of ${(n-1)^{th}}$ and ${n^{th}}$ transaction from $B$ to $A$, we need to scrutinize the number of transaction arrivals on edge $e_{AB}$ and the distribution of their total value. Formally, we provide the following theorem on the distribution of waiting time. 
% {\color{blue}
\begin{theorem}
The distribution of the waiting time $\Phi(v_k, u, t)$ is expressed as an iterative equation:
\begin{align}
\label{eq:third model}
           &\Phi(v_k,u,{t})=\sum\limits_{n=1}^{\infty}\sum\limits_{j=n}^{\infty}e^{-\lambda_2{t}}\frac{(\lambda_2{t})^{j}}{j!}(\varphi(r,n)-\varphi(r,n-1));\\
           &\varphi(r,n)=\sum\limits_{m=0}^{\infty}p_m\int_{0}^{\infty}q_n dG^{m^*}(z)+\varphi(r,1);\\
               &\varphi(r,1)=\sum\limits_{m=0}^{\infty}p_m\int\limits_{0}^{\infty}[1-G(r+z)]d(G^{m^*}(z));\\
           &q_n=\int_{0}^{u+z}\varphi(u+z-x,n-1)dG(x)\\
    &p_m=\frac{\lambda_1^m \lambda_2}{{(\lambda_1+\lambda_2)}^{m+1}};\\                      &r=v_k-u.
    \vspace{-1cm}
\end{align}
\end{theorem}
where $G(x)$ is the probability distribution function of $x$, $G^{m^*}(z)$ represents the convolution of $G(v_1)$ by $m$ times.
\noindent\textbf{Proof:} Our proof is carried out in two steps. 

\noindent\textit{Step 1: Equivalence between $\Phi(v_k,u,{t})$ and $\varphi(r,n)$}. 
Define $U(m)$ the capacity of edge $e_{AB}$ when there are $m$ transaction arrivals on edge $e_{BA}$. Define $T_k$ the inter-arrival time of two consecutive arrivals, $X_{k-1}$ and 
$X_k$, on edge $e_{AB}$. Then, the probability of the first passage event upon the arrival of $Y_m$ is given by:
\begin{align}
 &   U(m)=\sum\limits_{i=0}^{m}{v_{2i}}-\sum\limits_{i=0}^{N_1(\sum\limits_{k=1}^{m}T_k)}{v_{1i}}\\% +C_A^B(t_k)-V,
    &\varphi(r,n)=\textbf{Pr}\{U(m)\geq r, \exists m\leq n\}.
\end{align}
The first passage event is the union of mutually exclusive events that the first passage happens at the $n^{th}$ transaction arrival on edge $e_{BA}$. Accordingly, the waiting time distribution is expressed as 
\begin{align}
    \Phi(v_k,u,t)=\sum\limits_{n=1}^{\infty}\textbf{Pr}\{\sum\limits_{j=1}^{n}T_j<t|\lambda_2\}(\varphi(r,n)-\varphi(r,n-1)).
    \label{eq:phi}
\end{align}
Since the arrival processes are Poisson, the number of transaction arrivals on each edge is given by:
\begin{align}
   &\textbf{Pr}\{N_1(t)=m\} =\frac{(\lambda_1t)^me^{-\lambda_1t}}{m!},\\
      &\textbf{Pr}\{N_2(t)=m\} =\frac{(\lambda_2t)^me^{-\lambda_2t}}{m!}.
\end{align}
The inter-arrival time of transactions obeys the memoryless exponential distribution so that we can write down the sum of $n$  random variables in the form of Erlang distribution
\begin{align}
   & \textbf{Pr}\{T_n\leq t|\lambda\}=1-e^{-\lambda t}
    \label{eq:intervaln}\\
   & \textbf{Pr}\{\sum\limits_{i=1}^{n}T_i\leq t|\lambda_2\}=\sum\limits_{j=n}^{\infty}e^{-\lambda_2t}\frac{(\lambda_2t)^{j}}{j!}.
\end{align}

\noindent\textit{Step 2: Calculation of $\varphi(r,n)$}.
According to the full probability formula, we can obtain 
\begin{align}
&F_1=\textbf{Pr}\{U(m)\geq r,\exists m\leq n| T_1=t, \sum\limits_{i=1}^{y}v_{1i}=z, v_{21}=x\}\\\nonumber
&\varphi(r,n)=\textbf{Pr}(N_1(t)=y)\int\limits_{0}^{\infty}\textbf{Pr}(T_1=t)\int\limits_{0}^{\infty}\textbf{Pr}(\sum\limits_{i=1}^{y}v_{1i}=z)\\\nonumber
&\int\limits_{0}^{\infty}F_1d(G(x))dzdt\\\nonumber
&=\sum\limits_{y=0}^{\infty}\frac{(\lambda_1t)^y}{y!}e^{-\lambda_1t}\int\limits_{0}^{\infty}\lambda_2e^{-\lambda_2t}dt\int\limits_{0}^{\infty}\textbf{Pr}(\sum\limits_{i=1}^{y}v_{1i}=z)\int\limits_{0}^{\infty}F_1d(G(x))dz\\\nonumber
&=\sum\limits_{y=0}^{\infty}\frac{\lambda_2\lambda_1^y}{(\lambda_1+\lambda_2)^{y+1}}\int\limits_{0}^{\infty}\int\limits_{0}^{\infty}F_1d(G(x))d(G^{y^*}(z))
\end{align}
where $G^{y^*}(z)$ represents the convolution of $G(v_1)$ by $y$ times. We next derive the iterative formula as below:
\begin{itemize}
    \item $n=1$:
    $\int\limits_{0}^{\infty}F_1d(G(x))=\int_{r+z}^{\infty}d(G(x))=1-G(r+z)$;
    \item $n>1$: 
    $\int\limits_{0}^{\infty}F_1d(G(x))=1-G(r+z)+\int\limits_{0}^{r+z}\varphi(r+z-x,n-1)d(G(x))$.
\end{itemize}
Therefore, we can obtain
\begin{align}
    \varphi(r,n)=&\sum\limits_{y=0}^{\infty}\frac{\lambda_2\lambda_1^y}{(\lambda_1+\lambda_2)^{y+1}}
    \int\limits_{0}^{\infty}[1-G(r+z)+
    \\\nonumber&\int\limits_{0}^{r+z}\varphi(r+z-x,n-1)d(G(x))]d(G^{y^*}(z))
\end{align}
\begin{align}
    \varphi(r,1)=&\sum\limits_{y=0}^{\infty}\frac{\lambda_2\lambda_1^y}{(\lambda_1+\lambda_2)^{y+1}}\int\limits_{0}^{\infty}[1-G(r+z)]d(G^{y^*}(z))
\end{align}
Through Eq.\eqref{eq:phi}, the probability distribution of waiting time is obtained. See appendix for more details.

	\begin{table*}[h]
% 		\vspace{-0.2cm}
		\centering  
		
		\caption{The benefit of individual transaction in a two-player coalition.}
		\resizebox{0.8\textwidth}{!}{
		\begin{tabular}{|c|c|c|c|c|c|c|}
			\hline  
$S=\{X_0,X_1\}$&$\psi(S)=9.68$&$\psi(\{X_2\})=0$&$\psi(X_3)=0$ &   $\phi_{X_0}=4.84$ & $\phi_{X_1}=4.84$\\
\hline
$S=\{X_0,X_2\}$&$\psi(S)=1.2$ &$\psi(\{X_1\})=-0.8$&$\psi(\{X_3\})=0$& $\phi_{X_0}=0.6$ & $\phi_{X_2}=0.6$\\
\hline
$S=\{X_0,X_3\}$&$\psi(S)=0$ &$\psi(\{X_1\})=0$&$\psi(\{X_2\})=0$ &$\phi_{X_0}=0$ & $\phi_{X_3}=0$\\
			\hline 
$S=\{X_1,X_2\}$ & $\psi(S)=0.4$ &$\psi(\{X_0\})=0$&$\psi(\{X_3\})=0$&  $\phi_{X_1}=0.2$ & $\phi_{X_2}=0.2$\\
			\hline
$S=\{X_1,X_3\}$& $\psi(S)=0$ &$\psi(\{X_0\})=0$&$\psi(\{X_2\})=0$ &$\phi_{X_1}=0$ & $\phi_{X_3}=0$\\
			\hline
	$S=\{X_2,X_3\}$& $\psi(S)=0$&$\psi(\{X_0\})=0$&$\psi(\{X_1\})=0$&$\phi_{X_2}=0$ & $\phi_{X_3}=0$ \\
			\hline
		\end{tabular}}
		\label{table:shapely value result two}
% 		\vspace{-0.2cm}
	\end{table*}

	\begin{table*}[h]
% 		\vspace{-0.2cm}
		\centering  
		\caption{The benefit of individual transaction in a three-player coalition.}
			\resizebox{0.8\textwidth}{!}{
		\begin{tabular}{|c|c|c|c|c|c|c|}
			\hline  
$S=\{X_0,X_1,X_2\}$&$\psi(S)=10.08$ &$\psi(\{X_3\})=0$& $\phi_{X_0}=5.04$ & $\phi_{X_1}=4.64$ & $\phi_{X_2}=0.4$\\
\hline
$S=\{X_0,X_1,X_3\}$&$\psi(S)=9.68$ &$\psi(\{X_2\})=0$& $\phi_{X_0}=4.84$ & $\phi_{X_1}=4.84$ & $\phi_{X_3}=0$\\
\hline
$S=\{X_0,X_2,X_3\}$& $\psi(S)=1.2$ &$\psi(\{X_1\})=-0.8$ & $\phi_{X_0}=0.6$ & $\phi_{X_2}=0.6$ & $\phi_{X_3}=0$\\
			\hline 
$S=\{X_1,X_2,X_3\}$& $\psi(S)=0.4$ &$\psi(\{X_0\})=0$& $\phi_{X_1}=0.2$ & $\phi_{X_2}=0.2$ & $\phi_{X_3}=0$\\
			\hline
		\end{tabular}}
		\label{table:shapely value result three}
% 		\vspace{-0.4cm}
	\end{table*}

\section{Experimental Study}
In this section, we will show the experiment results of the coalitional mechanism and the stochastic waiting time. 
% \subsection{Experiment metrics}
% {\color{red} The second part can solve the problem the underutilization of the network.}%网络利用率

\subsection{overall result for coalitional mechanism}
\begin{figure}[h]
	\vspace{-0.3cm}
	\begin{minipage}[h]{0.5\textwidth}
		%\vspace{0cm}
		\centering
		\includegraphics[width=0.7\textwidth]{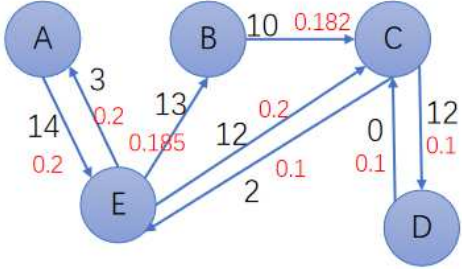}
		\caption{Topology of PCN in Shapley value case.}
		\label{fig:shapely value}
		%\vspace{0cm}
	\end{minipage}
	% 	\vspace{-0.5cm}
% 	\begin{minipage}[h]{0.24\textwidth}
% 		%\vspace{0cm}
% 		\centering
% 		%	\setlength\abovecaptionskip{-0.5pt}
% 		%	\setlength\belowcaptionskip{-1pt}
% 		\includegraphics[width=1\textwidth]{picture/shapely value2.PNG}
% 		\caption{Topology 2 of PCN in shapely value case.}
% 		\label{fig:shapely value2}
% 		%\vspace{0cm}
% 	\end{minipage}
\vspace{-0.3cm}
\end{figure}
First, we will use a specific PCN graph to study the coalitional mechanism. The initial topology of payment channel network is shown in Fig.\ref{fig:shapely value}. The integer represents the current capacity, decimal represents the charge rate of the channel. The basic routing fee for all channels is set to $0$ and $\xi$ is set to $10$. There are a set of transactions $\textbf{X}=\{X_0,X_1,X_2,X_3\}$, where $X_0=(D,A,1,0), X_1=(E,D,8,1), X_2=(A,C,6,2), X_3=(A,E,0.1,3)$. If each transaction chooses not to cooperate with others (in-order transfers upon their arrival times), the routing cost of each transaction is $f(X_0,\mathcal{G}_0)=\xi,  f(X_1,\mathcal{G}_1)=1.6, f(X_2,\mathcal{G}_2)=2.4, f(X_3,\mathcal{G}_3)=0$ respectively. Accordingly, the benefit for each user is $0$. If all the transactions choose to cooperate, the optimal order is $X_2, X_1, X_0, X_3$. The corresponding routing fee is $f(X_2,\mathcal{G}_0)=1.2,  f(X_1,\mathcal{G}_1)=2.4,  f(X_0,\mathcal{G}_2)=0.32,  f(X_3,\mathcal{G}_3)=0$. The benefit for grand coalition is $\psi(X)=10.08$. We can 
compute the benefits for the individual transaction.

The first row of Table \ref{table:shapely value result one} shows everyone's benefit if no one chooses to cooperate, and the second row shows that if a grand coalition is formed, each user's benefit will be distributed through Shapley value. After reordering, the routing fee decrease 72\% and the success rate of the number of transactions increase 25\%.
	\begin{table}[h]
% 		\vspace{-0.2cm}
		\centering  
		\caption{Benefit re-distribution of FCFS and coalition.}
		\begin{tabular}{|c|c|c|c|}
			\hline  
    $ \psi(\{X_0\})=0 $& $\psi(\{X_1\})=0$ & $\psi(\{X_2\})=0$ & $\psi(\{X_3\})=0$\\
			\hline 
	 $\phi_{X_0}=5.04$ & $\phi_{X_1}=4.64$ & $\phi_{X_2}=0.4$ & $\phi_{X_3}=0$\\
\hline		
		\end{tabular}
		\label{table:shapely value result one}
		\vspace{-0.4cm}
	\end{table}

If two players choose to cooperate with each other, the benefits are shown in Table \ref{table:shapely value result two}.	If there are there players choose to join in the coalition, the benefit is shown in Table \ref{table:shapely value result three}.

The result shows under this topology, the worth function of coalition and the re-distribution function fully meet the properties desired. 
% For $\psi(S)$, it shows:
% \begin{itemize}
%     \item Cohesive: The second to fourth columns of each row in table \ref{table:shapely value result two} and the second to three columns of each row in table \ref{table:shapely value result three} represents each partition of $X$. The sum of these columns in each line is less than $\psi(X)$.
%     \item Weak superadditivity: Table \ref{table:shapely value result one} table \ref{table:shapely value result two} and table \ref{table:shapely value result three} show the benefits of one player, two players and three players in the coalition, respectively. Its shows $\psi(S\bigcup X_i)\geq \psi(S)$ in each case.
% \end{itemize}
% For re-distribution function $\phi_i(S)$, it shows:
% \begin{itemize}
%     \item Rationality: According to the result shown in table \ref{table:shapely value result two} and table \ref{table:shapely value result three}, the benefit is greater or equal than non-cooperation for each player in coalition. For each coalition, the benefit is fully distributed to the players in coalition.
%     \item Balanced contribution: 
% \end{itemize}

\subsection{Success ratio of successful transactions in PCNs.}
We will show the performance of coalitional mechanism under different network parameters. The performance metric is success rate including success rate of transaction number and transaction value in PCN. Success rate of transaction number is the number of transactions which is processed successfully in PCN over the number of total transactions. Success rate of transaction value is the total amount of payments which is processed successfully in PCN over the total amount of payments generated. The transaction fee of trading on the public chain is much greater than that of trading on the PCN. Therefore, we ignore routing fees on PCN and focus on how many and how much of transactions can be successfully processed in PCNs through our mechanism. Also, we study the performance under different conditions, including different graph size and different number of transactions.
50 qualified network typologies are generated randomly, 50 sets of transactions are generated under each network topology, and each set of transactions includes some transactions. We compare the maximum and the minimum number and value of transactions which are processed successfully in PCNs and calculate the average for all sets of transactions under all topology as the result.

In our typologies, the number of nodes is set to 15, the number of edges is set to 70, the capacity of channel is randomly selected in the range of 10 coins and 15 coins. Considering that most of the transaction value is relatively small, we randomly select the transaction value in the range of 1 coin to 12 coins. Each node can choose its own charging standard, and we choose it randomly from 1\% to 5\%. Base routing fee is $0$ for all channels. Each set of transactions includes 4 transactions.

% \subsection{Performance under different nodes.}
\begin{figure*}[h]
% 	\vspace{-0.5cm}
	\begin{minipage}{0.24\textwidth}
		%\vspace{0cm}
		\centering
	\includegraphics[width=1\textwidth]{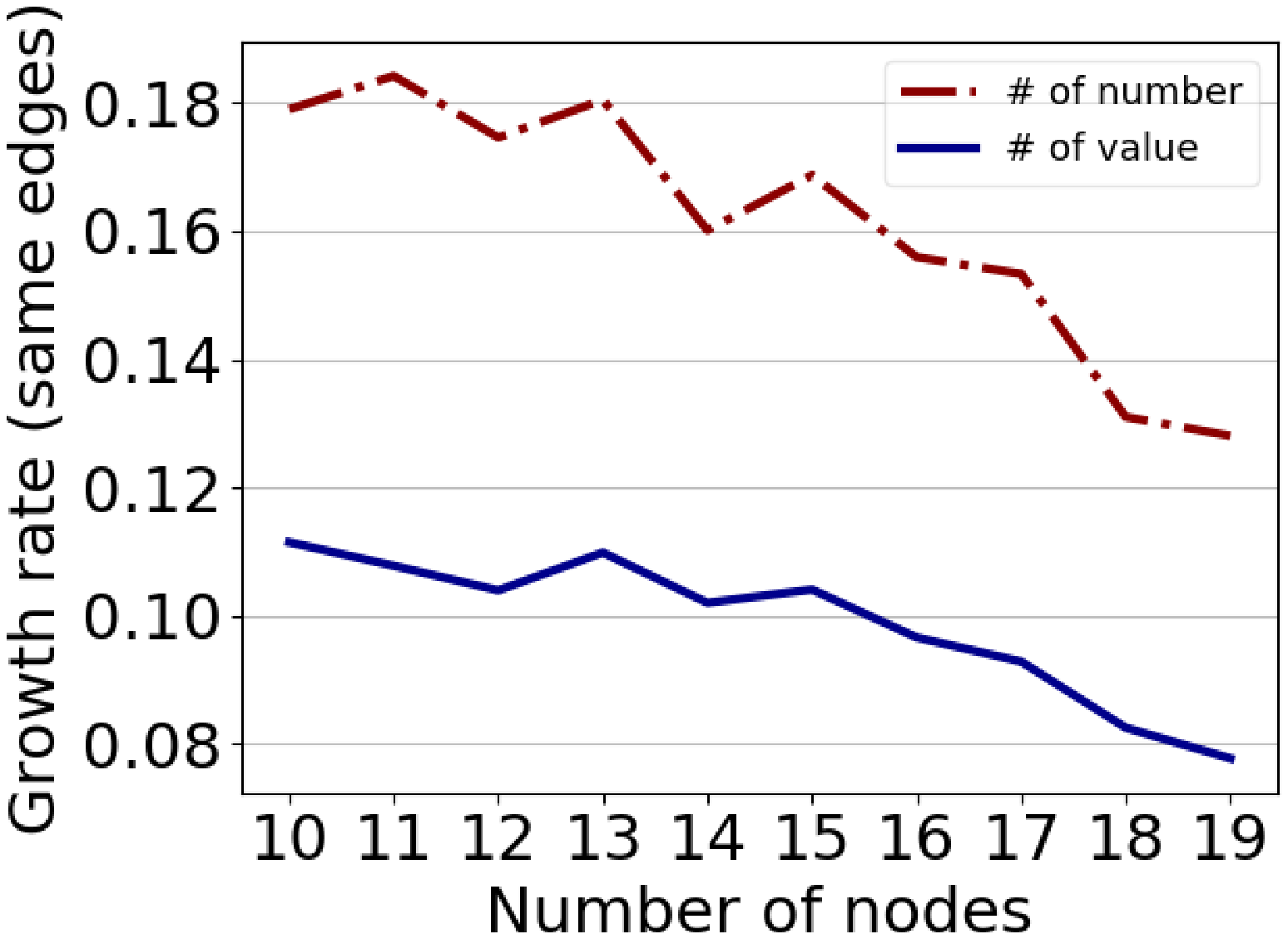}
		\caption{Growth rate of successful transactions in PCN w.r.t. number of nodes.}
		\label{fig:success ratio nodes}
	\end{minipage}
	\begin{minipage}{0.24\textwidth}
		%\vspace{0cm}
		\centering
	\includegraphics[width=1\textwidth]{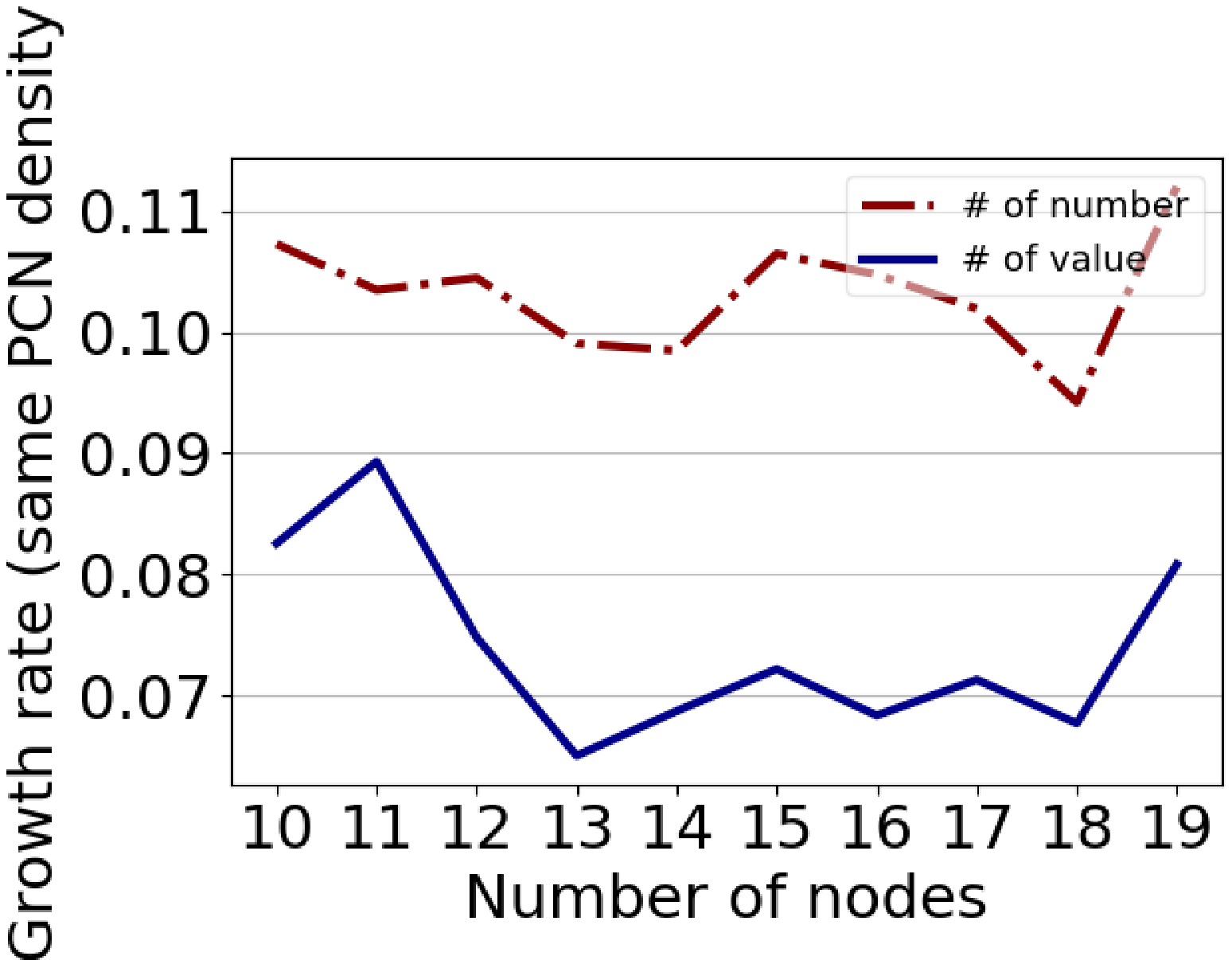}
		\caption{Growth rate of successful transactions in PCN w.r.t. number of nodes.}
		\label{fig:success ratio nodes same density}
		%\vspace{0cm}
	\end{minipage}
	\begin{minipage}{0.24\textwidth}
		%\vspace{0cm}
		\centering
	\includegraphics[width=1\textwidth]{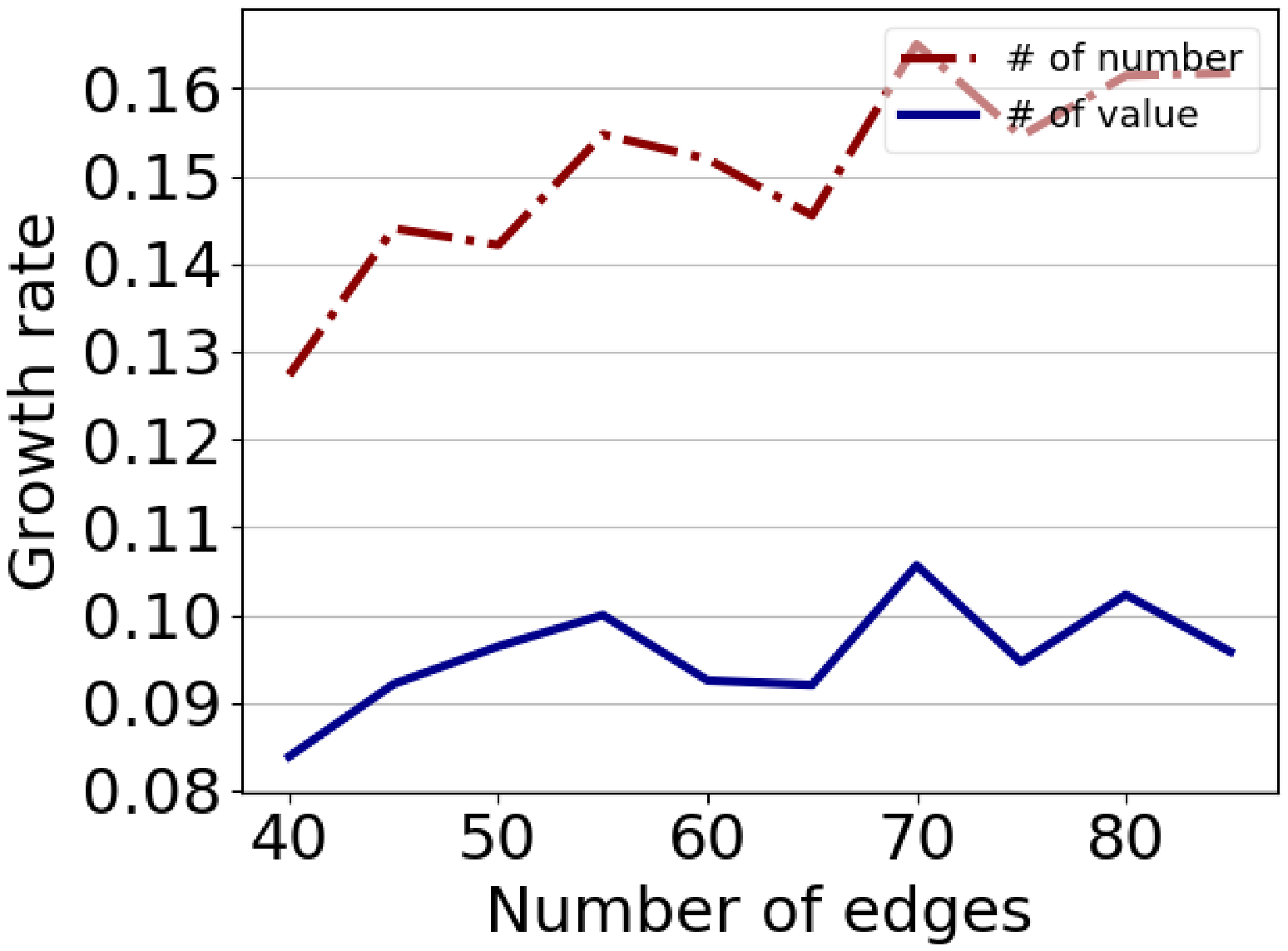}
		\caption{Growth rate of successful transactions in PCN w.r.t. number of edges.}
		\label{fig:success ratio edges}
		%\vspace{0cm}
	\end{minipage}
		\begin{minipage}{0.24\textwidth}
		%\vspace{0cm}
		\centering
	\includegraphics[width=1\textwidth]{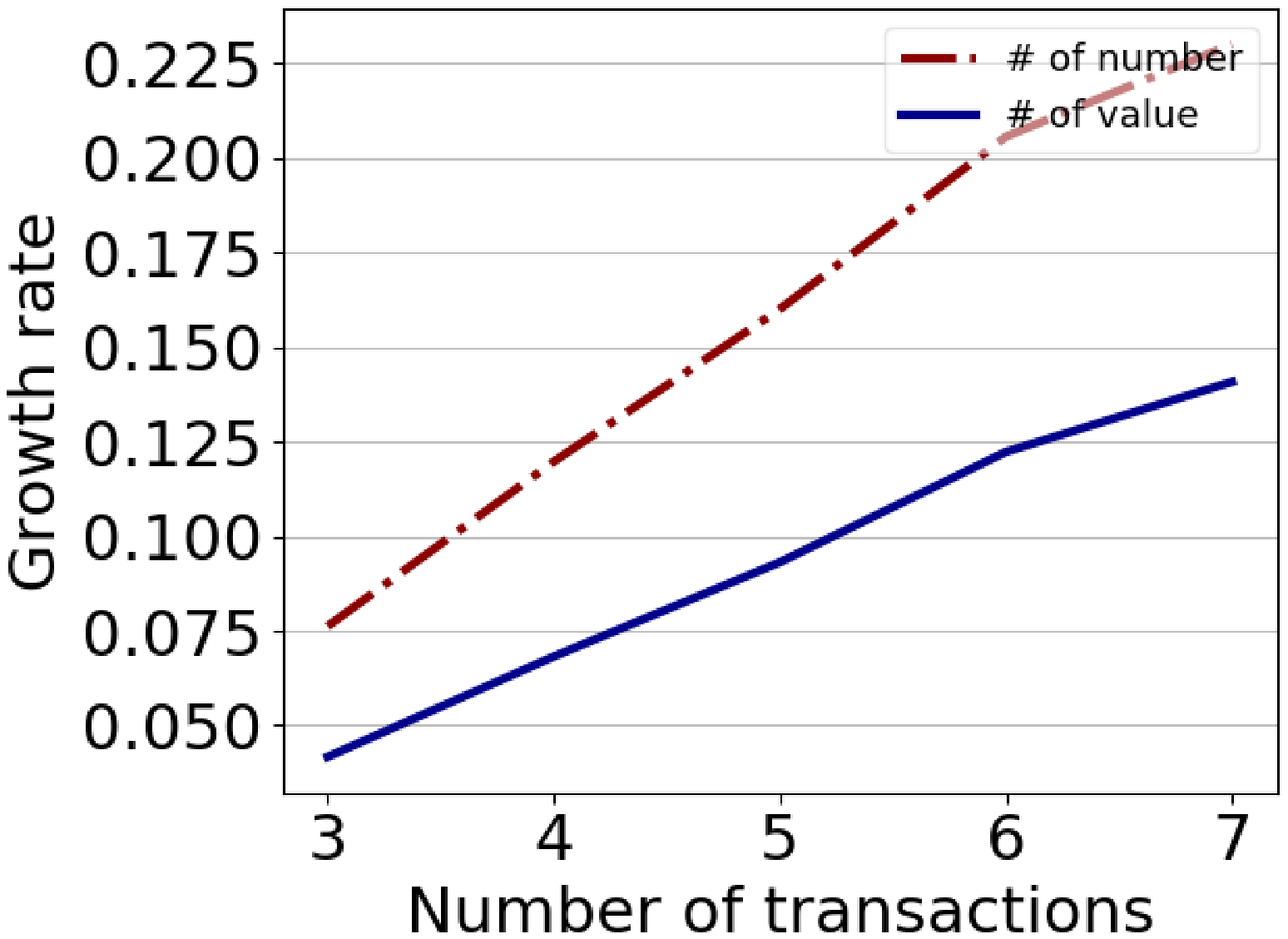}
		\caption{Growth rate of successful transactions in PCN w.r.t. number of transactions.}
		\label{fig:success ratio tx}
		%\vspace{0cm}
	\end{minipage}
\end{figure*}
Fig.\ref{fig:success ratio nodes} shows the result when the nodes change. With the increase of the number of nodes, the growth rates of successful transaction number and transaction value in PCN shows a downward trend. But the growth rate of successful transaction number is still more than 12\%, the growth rate of successful transaction value is more than 7\% when the nodes of PCN increases to 19. {The results manifest that the number and value of successful transactions of our proposed mechanism are significantly improved in the case of different number of nodes.} {In the case of the same number of edges, increasing the number of nodes (reducing network density) will lead to a decrease in the growth rate of the number and the amount of transactions which are processed successfully.} 

We want to explore whether the network density is one of the factors affecting the growth rate. Let us fix the network density and change the number of nodes. In PCN, If the direction of the capacity is not considered, the channel is actually undirected. Therefore, the density of PCN should be the density of undirected graph. The density\cite{ref:density} of undirected graph which contains $j$ nodes and $k$ edges is defined as:
\begin{displaymath}
\rho=\frac{k}{\frac{1}{2}j(j-1)}
\end{displaymath}
where the value of $\rho$ ranges from $0$ to $1$. The higher the value is, the denser the network is.
Fig.\ref{fig:success ratio nodes same density} shows the performance result of changing the number of nodes when $\rho =0.3$. The growth rate is relatively stable with the increase in the number of nodes. This proves success rate is relatively stable with the same density, regardless of the number or the value of successful transactions in PCN.

%edge:    node_num=15 c_range=[10,15] weight_range=[1,5] V_range=[1,12] tx_num=6
The performance of coalitional mechanism under different number of edges is researched.
Fig.\ref{fig:success ratio edges} shows in the case of the same number of nodes, growth rate of transaction number and transaction value both increase with the increase of the number of edges.

Number of transactions is also an important factor of performance for coalitional mechanism. In the above experiments, the number of transactions is set to $4$. The number of transactions per unit time is much more than $4$ in real system. Therefore we want to explore how the performance of the mechanism will change with the increase of the number of transactions under the same network size and related charges. Fig.\ref{fig:success ratio tx} shows with the increase of the number of transactions, the growth rate of the number and the value of successful transactions in PCN increases. {This means although the number of transactions in experiment is set to $4$, it still proves the effectiveness of our proposed mechanism. Because in real system, the more transactions, the more the number and value of successful transactions in PCN will increase.}

\subsection{Experiment results of stochastic waiting time.}
We assume that the transaction size obeys the exponential distribution with the parameter 2 coin, the transaction from A to B is a Poisson process with rate 1, and from B to A is a Poisson process with rate 2. The transaction amount to be transmitted is 10 coins, and the current capacity is 9 coins.
\begin{figure}[h]
% 	\vspace{-0.5cm}
	\begin{minipage}[h]{0.5\textwidth}
		%\vspace{0cm}
		\centering
		\includegraphics[width=0.7\textwidth]{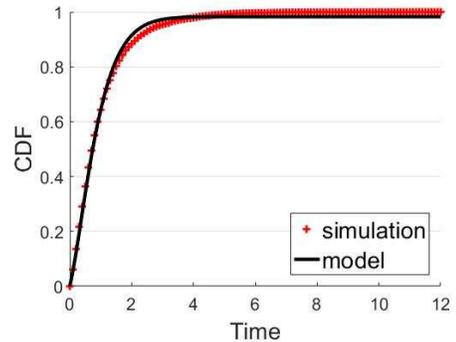}
		\caption{Results of model and simulation.}
		\label{fig:model and simulation}
		%\vspace{0cm}
	\end{minipage}
% \vspace{-0.4cm}
\end{figure}

We validate Eq\ref{eq:third model} and Fig.\ref{fig:model and simulation} show the result. In Fig.\ref{fig:model and simulation}, the points represent the simulation results, the black line represents the model result. Horizontal axis is the time, and the vertical axis is cumulative probability function of $\widehat{t}$. The simulation results are basically matched with the theoretical results. The error is mainly caused by numerical integration. For the integral calculation of the original function, we use use the numerical integration method instead of the Newton-Leibniz formula. Considering the running time of program, it is necessary to select the appropriate numerical integration step, which will introduce errors.
% \begin{figure}[h]
% % 	\vspace{-0.5cm}
% 	\begin{minipage}[h]{0.5\textwidth}
% 		%\vspace{0cm}
% 		\centering
% 		%	\setlength\abovecaptionskip{-0.5pt}
% 		%	\setlength\belowcaptionskip{-1pt}
% 		\includegraphics[width=0.8\textwidth]{picture/success_ratio/model_C.png}
% 		\caption{Result of probability w/ $(V-C)$.}
% 		\label{fig:model v-c}
% 		%\vspace{0cm}
% 	\end{minipage}
% % \vspace{-0.5cm}
% \end{figure}

\begin{figure}[h]
% 	\vspace{-0.5cm}
	\begin{minipage}[h]{0.5\textwidth}
		\vspace{-0.4cm}
		\centering
		\includegraphics[width=0.7\textwidth]{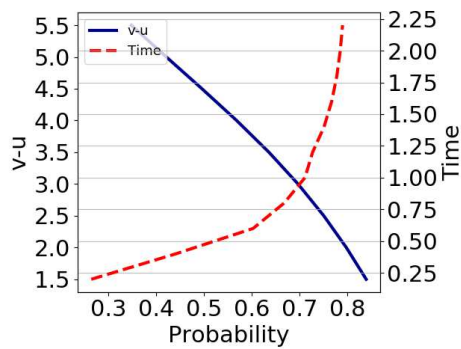}
		\caption{Result of probability w.r.t. $t$ and probability w.r.t. $v-u$.}
		\label{fig:model}
		%\vspace{0cm}
	\end{minipage}
% \vspace{-0.4cm}
\end{figure}

Fig. \ref{fig:model} shows the results of $\Phi(v,u,t)$ with different parameters. Here, $x-$coordinate is the probability result of $\Phi(v,u,t)$ , and left $y-$axis is the difference between the amount to be transferred and the current capacity $(v-u)$, right $y-$axis is different limited time $(t)$. 
The dotted line shows that with the increase of the transmission amount, the probability of reaching the requirement before $3$ decreases continuously. The solid line illustrates with the increase of limited time, the probability increase from $0.25$ to $0.8$. 
\section{Conclusion}
In this paper, we study how to find the payment channel routing with minimum cost in payment channel network. We focus on this problem from the point of view of whether we can get lower transaction fees by waiting patiently instead of pursuing maximum efficiency.
A periodic transaction processing scheme is designed and it re-orders the transactions to get an minimum routing fee. The benefit is re-distributed by Shapley value which ensures the fairness. 
Also, The capacity dynamics of a payment channel is captured by a transient queuing model and the waiting time distribution of a strategic transaction is obtained.

\end{document}